\documentclass[12pt]{iopart}
\usepackage[latin1]{inputenc}
\usepackage[dvips]{epsfig}
\usepackage{graphicx}
\usepackage{times}

\begin{document}

\title{Biomimetic membranes of lipid-peptide model systems prepared on solid support}
\author{Chenghao Li, Doru Constantin and Tim Salditt}
\address{Institut f\"{u}r R\"{o}ntgenphysik, Univ. G\"{o}ttingen, Geiststra{\ss}e 11, D-37037 G\"{o}ttingen, Germany}

\ead{tsaldit@gwdg.de}

\begin{abstract}
The structure of membrane-active peptides and their interaction with lipid bilayers can be studied in oriented
lipid membranes deposited on solid substrates. Such systems are desirable for a number of surface-sensitive
techniques. Here we focus on structural characterization by X-ray and neutron reflectivity and give account
of recent progress in sample preparation and measurements. We show that the degree of mesoscopic disorder
in the films can significantly influence the scattering curves. Static defects should be minimized by
optimization of the preparation techniques and their presence must be taken into account in the modelling.
Examples are given for alamethicin and magainin in bilayers of different phosphocholines.

\end{abstract}




\vspace{1em}
\noindent \textit{J. Phys.: Condens. Matter} \textbf{16}(26), p. S2439 (2004).\\
DOI: 10.1088/0953-8984/16/26/017


\section{Introduction}

Compared to isotropic bulk solutions, biomimetic membranes deposited on solid surfaces offer a number of
advantages for structural studies~: surface characterization techniques can be applied to study such
systems, different symmetry axes of the system can be separated, and macromolecular conformation of
proteins or peptides can possibly be probed in and at the bilayer. From a technological point of view,
solid surfaces may be used in future to manipulate or detect interactions in the biomolecular films
deposited on top of them. Biomimetic interfaces and bio-functional surfaces are therefore an active field
of interdisciplinary research. To this end, the preparation of well-defined, homogeneous and structurally
intact membrane systems on solid support is an important problem, involving fundamental physical questions,
for example  related to wetting behaviour \cite{Gallice:2002}, thermal stability
\cite{Vogel:2000,Pabst:2002,PozoNavas:2003}, or defects typical for smectic liquid crystalline (LC) films.
A particularly simple and low-cost approach to preparing oriented lipid membranes is to spread or to
spin-coat a solution of co-dissolved lipids and peptides onto solid surfaces like silicon, glass, quartz or
other flat surfaces \cite{Seul:1990,Mennicke:2002}.

\begin{figure}
\begin{center}
\epsfig{file=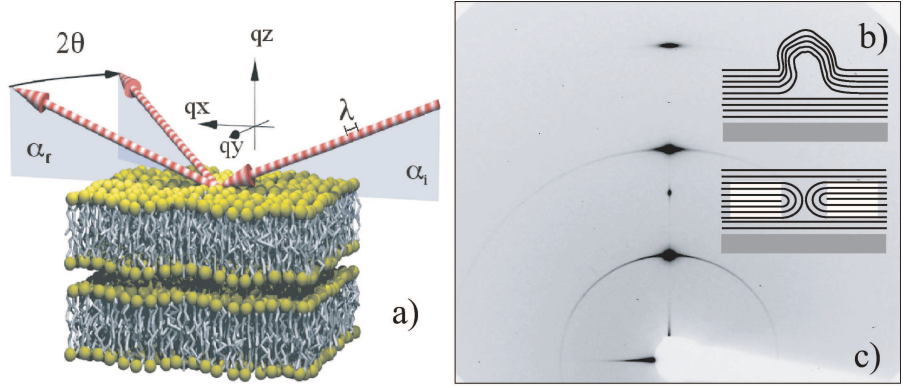,width=12cm}
\end{center}
\caption{\label{fig:fig1} a) Sketch of an interface-sensitive scattering geometry, showing the incidence
angle $\alpha_{i}$, the reflection angle $\alpha_{f}$ and the out-of-plane angle $2\theta$. While the
reflected beam carries the information on the vertical sample structure ({\em i.e.} the bilayer density
profile), the diffuse scattering reveals thermal fluctuations and/or static defects. b) Possible structure
for lamellar phase defects~: multilamellar vesicles (top) and giant dislocations. c) Structure of the
reciprocal space, illustrated by a CCD image taken on the ID1 beamline at ESRF. The sample is a 1/30
mixture of Alamethicin/DMPC, immersed in a $31 \%$ PEG solution (molecular weight $20 000$), with 100 mM
NaCl; see also figure \ref{fig:Ala_DMPC}. Three Bragg sheets are visible as extended spots, as well as the
sharp rings due to the presence of defects.}
\end{figure}

The structure of biomimetic multilamellar lipid membranes deposited on solid surfaces can be studied at
high resolution by modern surface-sensitive scattering techniques, using synchrotron based X-rays or
neutrons as a probe. These techniques offer a novel approach to investigating the structure of lipid
bilayer systems, both with and without additional membrane-active molecules such as amphiphilic peptides or
membrane proteins. A high degree of orientation (low mosaicity) makes possible a precise distinction
between the scattering vector component normal $q_z$ and parallel $q_{\parallel}$ to the bilayer, opening
up a way to study questions associated with the lateral structure of the bilayers. Most diffraction studies
of aligned lipid films have been carried out in the low temperature gel phase \cite{Katsaras:2000}. Here we
concentrate on the physiologically more relevant liquid $L_\alpha$ phase.

Figure 1(a) presents the diagram of an interface-sensitive scattering experiment~: the incoming beam makes
an angle $\alpha_i$ with the plane of the bilayers. The outgoing beam makes an angle $\alpha_f$ and it can
also deviate by an angle $2 \theta$ from the plane of incidence (defined by the incident beam and the
bilayer normal, taken in the following as the $z$ axis.) Multilayer systems, however well oriented on the
average, often exhibit defects such as those depicted in figure 1(b), in which the normal to the bilayers
makes a large angle with the average direction $\vec{z}$. Such defects (which can be minimized by careful
preparation techniques or by long annealing times) are responsible for the sharp rings appearing at
$q_n = 2\pi n/d$, as in the CCD image of figure 1(c), intersecting the three diffuse Bragg sheets. The
bright spot between the first and second Bragg sheets is given by the specular beam ({\em i. e.} $\alpha_i
= \alpha_f$.)

A few years ago we started using aligned multilamellar membranes to study structural details of the
lipid-peptide interaction by X-ray and neutron reflectometry. The vertical density profile of the bilayers
$\rho(z)$ (averaged over the $xy$-plane) can be determined from least-square fitting of the specular
reflectivity curve, while diffuse scattering reflects the lateral inhomogeneities on mesoscopic length
scales, from a few nm up to a few $\mu$m. One aim of such studies is to deduce the peptide position and
conformation with respect to the bilayer, {\em i. e.} to distinguish between the inserted and adsorbed
states. In principle, the corresponding bilayer density profiles are different for the two conformations,
and careful measurements should be able to evidence this. However, the observed changes in the reflectivity
curves are often dominated by effects of the structure factor due to changes in the static or thermal
fluctuations induced by the peptides. This difficulty has so far prevented us from extracting reliable
structural parameters. In the present work, we give account of this problem and discuss different
preparation procedures of multilamellar solid-supported bilayers, as well as the structure of these films
on mesoscopic length scales. Importantly, this mesostructure has strong implications on the structural
characterization on the molecular scale. We then show how a proper choice of organic solvents can minimize
these secondary effects, leading to more homogeneous films amenable to quantitative structural analysis.

The particular peptides which we are interested in here belong to a family of innate host defense molecules
known as antimicrobial peptides, see reviews in \cite{Bechinger:1999,Matsuzaki:1999,Huang:2000}. Well known
examples are ceropins expressed in insects, or magainin, the first antimicrobial peptide discovered in
vertebrates. Magainin is expressed in the intestines and the skin of the frog {\it Xenopus laevis}. Mammals
also express antimicrobial peptides called defensins, which lyse microbes, probably by destroying the
integrity of their cell walls, but leave the plasma membranes of their hosts intact. Other examples of
similar antibiotic peptides are cytolytic to mammalian cells, like the well known alamethicin of the fungus
{\it Trichoderma viride}, or the honey bee venom melittin. Host-defense and cytolytic peptides are
amphiphilic polypeptides of typically 20 to 40 amino acid residues, with well-defined secondary structures
forming upon interaction with the lipid bilayer. It has been shown that antimicrobial peptides interact
directly with the microbial cell membranes, rather than with specific membrane proteins, subsequently
causing an increase in membrane permeability and cell lysis. Apart from the obvious significance in the
biological and pharmaceutical sciences, membrane-active peptides pose many interesting questions of
biomolecular self-assembly in the bilayer and can be regarded as a testing ground for concepts and methods
which may then be translated to more complex membrane protein systems. However, despite recent advances due
to the use of a large number of different techniques, most models of the functional interaction and
structure remain partially hypothetical and incomplete, and require in-depth structural characterization.
To this end, a refinement of scattering techniques including sample preparation, measurement and data
analysis is needed, and could complement and significantly extend the possibilities of present small angle
scattering experiments \cite{WhiteHristova:2000,Nagle:2000}.

After this introduction, section 2 presents the sample preparation procedure and some experimental aspects
of the reflectivity measurements. Section 3 is devoted to the analysis and modelling of reflectivity curves
obtained from multilamellar lipid bilayers. A popular method for determining the bilayer density profiles
$\rho(z)$ is the Fourier synthesis based on Bragg peak intensities. We investigate the validity of this
method by comparing the results to those obtained from data fitting over the full $q_z$-range. Section 4
discusses the effect of thermal fluctuations and static defects on the structure determination and
presents different experimental routes for the structural analysis of supported multilamellar stacks by
scattering. Examples using these methods are presented in section 5 on model systems composed of pure
lipids and antibiotic peptides in lipids. Finally, section 6 presents a comparative discussion and some
tentative conclusions.

\section{Sample preparation, sample environment, and reflectivity setup}

Highly oriented membranes are characterized by a very small mosaicity ({\em i. e.} distribution of the
bilayer normal vector), typically of the order of $0.01 {\,}^{\circ}$ or below, which is small compared to
the critical angle for total external reflection of X-rays or neutrons. Therefore, quantitative interface-sensitive X-ray or neutron scattering methods like specular and non-specular reflectivity, grazing
incidence diffraction, and reciprocal space mappings, see figure 1 and recent monographs
\cite{Tolan:1999,Als-Nielsen:2001}, can be applied. Films of lipid membranes can be prepared using the
classical procedure of spreading lipids (and peptides) from solution \cite{Seul:1990}, as free standing
films \cite{Smith:1988}, or by more recent schemes which allow a precise control of the total number of
bilayers $N$ by spin-coating the solutions \cite{Mennicke:2002}. Novel methods to prepare single or double
bilayers by Langmuir techniques have also been reported \cite{Fragneto:2001}.

\begin{table}
\caption{\label{table:solubility} Solubility of lipids and peptides in different solutions~: (1)
2-propanol, (2) TFE (2-2-2-trifluoroethanol), (3) methanol, (4) ethanol, (5) chloroform, (6) chloroform/methanol(1:1), (7) acetone,
(8) TFE/chloroform, (9) TFE/ethanol(1:1), (10) HFI/chloroform, (11) HFI (1,1,1,3,3,3-hexafluoro-2-propanol).}
\begin{tabular}{cccccccccccc}
\br {\tiny } & {\scriptsize 1} & {\scriptsize 2} & {\scriptsize 3} & {\scriptsize 4} &
{\scriptsize 5} & {\scriptsize 6} & {\scriptsize 7} &{\scriptsize 8} & {\scriptsize 9}&{\scriptsize 10} & {\scriptsize 11}\\

\mr {\scriptsize DLPC}     & {\tiny Y} & {\tiny Y} &  &  & {\tiny Y} &  &  &  &  & {\tiny Y} & {\tiny Y}\\
\hline {\scriptsize DMPC}  & {\tiny Y} & {\tiny Y} &  &  & {\tiny Y} &  &  & {\tiny Y} & {\tiny Y} & {\tiny Y} & {\tiny Y}\\
\hline {\scriptsize DOPC}  & {\tiny Y} &  &  &  &  &  &  &  &  & {\tiny Y} & {\tiny Y}\\
\hline {\scriptsize DPPC}  & {\tiny Y} & {\tiny Y} &  &  & {\tiny Y} &  &  &  &  & {\tiny Y} & {\tiny Y}\\
\hline {\scriptsize OPPC}  & {\tiny Y} & {\tiny N} &  &  & {\tiny Y} &  &  & {\tiny Y} & {\tiny Y} & {\tiny Y} & {\tiny Y}\\
\hline {\scriptsize DMPE}  &  & {\tiny N} &  &  & {\tiny Y ($<$ 1mg/ml)} &  &  & {\tiny Y (after 24 h)} &  & {\tiny Y} & {\tiny N}\\
\hline {\scriptsize POPE}  & & {\tiny N} &  &  & &  &  & {\tiny Y} &  & {\tiny Y} & {\tiny Y}\\
\hline {\scriptsize DMPG}  & & {\tiny N} &  &  & {\tiny Y ($<$ 1mg/ml)} &  &  & {\tiny Y} &  & {\tiny Y} & {\tiny Y}\\
\hline {\scriptsize POPS}  &  & {\tiny N} &  &  & {\tiny Y} &  &  & {\tiny Y (TFE/CH$_3$Cl$\geq 1/3$)} &  & {\tiny Y} & {\tiny N}\\
\hline {\scriptsize Alam.} & {\tiny Y} &  & {\tiny N} & {\tiny N} & {\tiny N} & {\tiny Y} & {\tiny N}  &  &  &  & \\
\hline {\scriptsize Mag.2} &  & {\tiny Y (about 1mg/ml)} &  &  & {\tiny N} &  &  & {\tiny Y} & {\tiny Y} & {\tiny Y} & {\tiny Y}\\
\hline {\scriptsize Gram. D}  & {\tiny Y} &  & {\tiny Y} & {\tiny Y} & {\tiny Y} &  & {\tiny Y} &  &  &  & \\
\br
\end{tabular}
\end{table}

\subsection{Sample preparation for the presented experiments} 1,2-dimyristoyl-sn-glycero-
3-phosphatidylcholine (DMPC), 1,2-oleoyl-palmitoyl-sn-glycero-3-phosphatidylcholine (OPPC),
1,2-dimyristoyl-sn-glycero-3-phosphoethanolamine (DMPE) were bought from Avanti Lipids, Alabama, and used
without further purification. The peptide magainin 2 amide ($GIGKFLHSAKKFGKAFVGEIMNS$) was obtained by
solid-phase peptide synthesis by B. Bechinger and coworkers \cite{Bechinger:1999}. Alamethicin
($XXPXAXAQXVXGLXPVXXEQ$) was bought from Sigma (product number: 05125). Multilamellar bilayers were
prepared on cleaned silicon or glass wafers by spreading from organic solution, similar to the procedure
first described by Seul and Sammon \cite{Seul:1990}. The challenge is to simultaneously meet the solvation
and wettability requirements. For sample deposition the substrates were cleaned by two 15 min cycles of
ultrasonic cleaning in methanol, followed by two 15 min cycles in ultrapure water (specific resistivity
$\ge$ 18M$\Omega$~cm, Millipore, Bedford, MA), and drying under nitrogen flow. Finally, they were rendered
hydrophilic by etching in a plasma cleaner (Harrick Scientific, NY) for 30 seconds. The lipid and peptide
components were co-dissolved in the desired proportions (molar ratio P/L) in (i) 2-propanol, (ii)
2-2-2-trifluoroethanol (TFE) and (iii) 1:1 Chloroform:TFE mixtures (see table \ref{table:solubility} for
solubility of lipid and peptide in some solvents), at total concentrations between 4 and 20 mg/ml,
depending on the total mass to be deposited. A drop of 0.1 ml was then carefully spread onto well-leveled
and cleaned Si (100) or glass substrates of typically $15\times 25 \, \mbox{mm}^2$ yielding average film
thicknesses of about $D\simeq 5-10 \, \mu$m. The spread solution was allowed to dry only very slowly to
prevent film rupture and dewetting. The films were then exposed to high vacuum over 24 hours in order to
remove all traces of solvent and subsequently rehydrated in a hydration chamber. In all cases, the
mosaicity was typically better than 0.02$^\circ$. To monitor sample quality, we used light microscopy in
DIC contrast (Zeiss Axioskop, objective: Neofluar 20x /0.5 DIC) to image the samples after deposition,
mainly in the dry state. Figure 2 shows DMPC samples of 15 mg/ml stock solutions (0.2 ml spread on $15
\times 25\, \mbox{mm}^2$ Si (100) in three different organic solvents, namely isopropanol, TFE, and 1:1
TFE:chloroform mixture. The representative images clearly show that the defect structure depends on the
solvent. In general, TFE:chloroform mixtures gave the best results, {\em i.e.} the smallest number of
defects.

\begin{figure}
\begin{center}
\epsfig{file=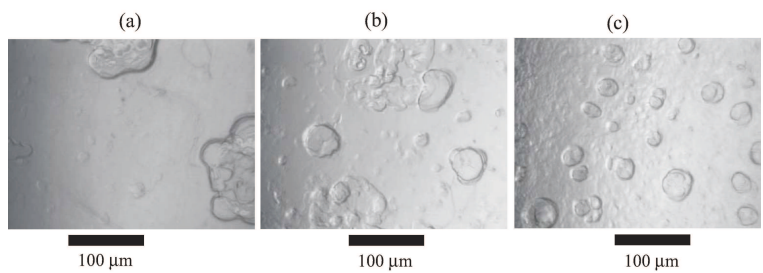,width=10cm}
\end{center}
\caption{\label{fig:microscopy}Light microscopy of pure DMPC samples, spread from (a) 1:1 mixtures of TFE
and chloroform, (b) TFE, (c) isopropanol, all at 15 mg/ml concentration.}
\end{figure}

\subsection{Sample environment} Sample environment for the control of temperature, humidity, and possibly other
parameters (osmotic pressure, electrical fields etc.) can generally be made compatible with the X-ray
experiments. Here, the sample chamber consisted of two stainless steel cylinders with kapton windows. The
chamber was cooled or heated by a flow of oil or 1:2 glycol:water mixtures from a temperature-controlled
reservoir (Julabo, Germany). The samples were mounted in an inner chamber with a water reservoir to keep
the relative humidity close to $100 \%$. The temperature was measured in most cases in the inner chamber by
a Pt100 sensor, showing a stability of better than $0.03$ K over several hours. A sensor for relative
humidity (HIH2610-003, Honeywell, Freeport IL) was additionally installed, but in most cases failed to give
reproducible results near $100 \%$ relative humidity. In most of our measurements, uncharged membranes
could not be swollen to their equilibrium periodicity $d_0$ in water vapor, even if the vapor was
(nominally) at $100 \%$ relative humidity. This phenomenon, long known as the vapor pressure paradox,
results from small temperature gradients in the sample chamber \cite{Nagle:1999}. In practice, we took the
membrane periodicity $d$ of pure DMPC as a control of the humidity at a given temperature and chamber
mounting. It is also possible to study solid-supported lipid films immersed in excess water
\cite{Vogel:2000}. This is of interest for two reasons~: Firstly, excess water warrants the physiologically
relevant condition of full hydration. Secondly, membrane-active molecules can be adsorbed directly from the
solution. However, films in excess water are unstable in the absence of osmotic agents (stressors). A
thermal unbinding transition was observed \cite{Vogel:2000,Pabst:2002}, from a substrate-bound,
multilamellar state at low temperatures to a state of freely dispersed bilayers in water at high
temperatures.  Unbinding can be suppressed (and the films thus stabilized) by adding an osmotic stressor to
the excess water. The control of the periodicity $d$ can be achieved by the use of excess polymer solutions
as osmotic stressors, and the equation of state can be determined \cite{Mennicke:2002,Brotons:2003}. For
charged systems, mixing of the bilayers and stressor polymers can be avoided by using polyelectrolytes of
the same charge as the lipids. In this work, we used samples immersed in water for two series of
DMPC/Alamethicin at different osmotic pressures, using as osmotic stressor solutions of (i) $31 \%$ PEG
solution (molecular weight $20 000$), with 100 mM NaCl, and (ii) $14.2 \%$ PEG at zero salt (see figure
\ref{fig:Ala_DMPC}).

\subsection{X-ray reflectivity} For this method, the incident beam with wave vector $\mathbf{k}_i$ has to be
collimated to less than a few hundredths of a degree and directed onto the sample at a glancing incidence
angle $\alpha_i$. The reflected intensity is then measured as a function of $\alpha_i$ under specular
conditions, {\em i.e.} at an exit angle $\alpha_f=\alpha_i$ and out-of-plane angle $2 \theta =0$, with the
wave vector of the exit beam denoted by $\mathbf{k}_f$. Thus, the momentum transfer of the elastic
scattering ${\bf q}=\mathbf{k}_f -\mathbf{k}_i$ is always along $q_z$, with the $z$-axis parallel to the
sample normal, see figures 1 and 2. Contrarily, moving the detector or sample to $\alpha_i \ne \alpha_f$
(diffuse or non-specular scattering) results in a component $q_\parallel$ parallel to the sample surface.

The reflectivity measurements presented here were carried out at three different experimental stations~:
(i) At a high resolution in-house rotating anode reflectometer, equipped with a channel-cut
$\mbox{Ge}(110)$ monochromator selecting the Cu K$_{\alpha 1}$ radiation, a $z$-axis diffractometer and a
standard NaI scintillation counter. (ii) At the bending magnet beamline D4 of HASYLAB/DESY using photon
energies of 20 keV. (iii) At the experimental station ID1 of ESRF Grenoble using photon energies of 19 keV.
Incidence and exit beams are defined by various slits. Typically, the reflectivity can be recorded over
seven to eight orders of magnitude, after correction for diffuse scattering background, as measured in an
offset scan. In order to get correct results when fitting the reflectivity, the correction for diffuse
background at higher angles is essential. Beyond background subtraction, the diffuse (non-specular)
scattering component contains valuable information on the lateral membrane structure on mesoscopic length
scales, in particular the height fluctuations as quantified by the height-height self- and
cross-correlation functions \cite{Salditt:1999}.

\section{Analysis of reflectivity curves from multilamellar membranes}

The analysis of X-ray and neutron reflectivity requires a very low mosaicity (narrow orientational
distribution of domains) as well as a flat substrate, allowing a clear separation between the specular and
non-specular scattering components. Two standard approaches are the fully dynamical Parratt algorithm
(taking into account multiple reflections) or the semi-kinematical reflectivity pioneered by Als-Nielsen
\cite{Als-Nielsen:2001}. In contrast to small-angle scattering, the observation of a region of total
external reflection and hence of the critical angle $\alpha_c$ allows the determination of the electronic
density profile on an absolute scale $[\mbox{e}^{-}/ \mbox{\AA}^3]$. Furthermore, since the full
$q_z$-range can be used for data analysis by fitting the reflectivity curve to a parameterized model of the
density profile \cite{Salditt:2002}, a reasonable resolution in $\rho(z)$ can also be reached for fully
hydrated systems. Compared to arbitrary interface profiles, the analysis is significantly simplified, since
the bilayer form factor is real-valued due to centro-symmetry and changes in sign are often accompanied by
an observable cusp in the (continuously measured) reflectivity curve. Alternatively, phasing of the Fourier
components can be performed by the so-called swelling method. Note that the advantage of full $q_z$-fitting
has also been demonstrated in bulk (SAXS) studies, see for example \cite{Pabst:2000}. In most published
studies of oriented bilayers, however, only the integrated Bragg peaks of the multilamellar samples are
used for data analysis, and the one-dimensional density profile $\rho(z)$ is computed by Fourier synthesis
using a discrete set of coefficients $f_n$ as described in \cite{Blaurock:1982,Katsaras:1995,Zhang:1994}
(see equation (\ref{eq.fc}) below).

\begin{figure}
\begin{center}
\epsfig{file=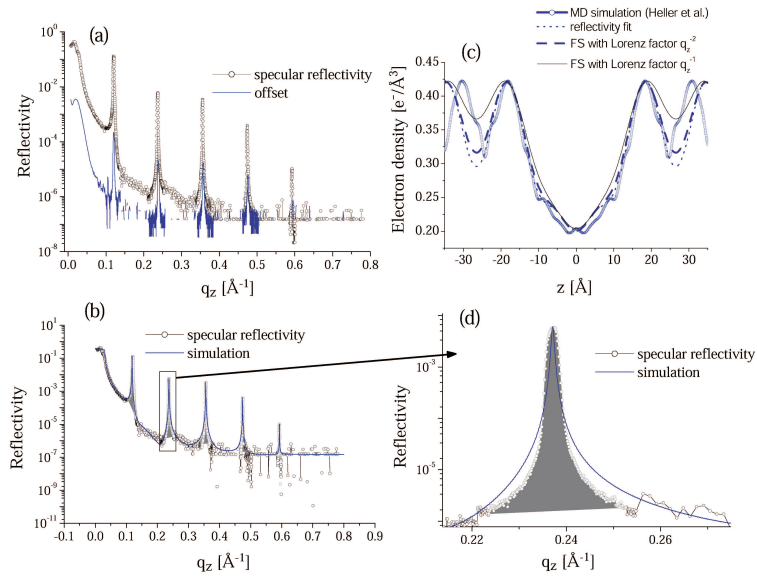,width=10cm}
\end{center}
\caption{\label{fig:pure_OPPC}(a) Reflectivity and offset scan (non-specular background) of multilamellar
OPPC membranes in the fluid $L_{\alpha}$ phase at partial hydration (T=45°C, and $d \simeq 53 \mbox{\AA}$),
(b) True specular component obtained by subtraction of the offset scan, fitted to a model with 5 free
Fourier components which define the electron density profile on an absolute scale (see text). The grey
areas indicate the integrated peak intensities, which are used to calculate the electron density profile by
Fourier synthesis, shown in (c) for the two alternative Lorentz factors,  $q^{-2}_z$(dashed line) and
$q^{-1}_z$ (thin solid line), along with the result of the full $q$-range fit (solid line in (b)), and the
profile calculated from MD simulations. (d) Zoom of the second Bragg peak of (b), showing the deficiencies
of the model in terms of resolution effects (see text).}
\end{figure}

We have recently developed a reflectivity model in the framework of semi-kinematical scattering theory, in
which both the structure factor of the stack and the bilayer form factor can be suitably chosen
\cite{Salditt:2002}, according to the given experimental resolution. This is possible since the lipid
bilayer density profile $\rho_{bl}(z)$ and the associated form factor $F(q_z)$ are parameterized by a
variable number $No$ of Fourier coefficients, where $No$ is adapted to the resolution of the measurement.
In contrast to conventional box models, the total number of parameters can thus be kept small, while still
fitting to reasonable density profiles. As a test example, the reflectivity of highly aligned multilamellar
OPPC membranes on solid substrates has been measured and analyzed \cite{Salditt:2002}. The resulting
density profile agrees remarkably well with the bilayer structure as obtained from published molecular
dynamics (MD) simulations \cite{Heller:1993}, (see figure \ref{fig:pure_OPPC}(b)). The starting point for
this treatment is the so-called master equation of reflectivity from a structured interface in
semi-kinematic approximation \cite{Als-Nielsen:2001}. For an interface perpendicular to the $z$ axis,
characterized by the (laterally averaged) scattering length density profile $\rho (z)$, (electron density
for X-rays) between a medium 1 (air or water) with scattering length density $\rho_1$ and a medium 2 (solid
substrate) with density $\rho_2$, the reflectivity reads~:
\begin{equation}
\label{eq.me} R (q_z)=R_F (q_z) \left| \Phi(q_z) \right|^2 = R_F
(q_z) ~\left| {\frac{1}{\Delta \rho_{12} } \int \frac{\partial
\rho(z)} {\partial z} e^{-\rmi q_zz}\,\rmd z\ }\right|^2 ~,
\end{equation}
where $R_F$ is the Fresnel reflectivity of the ideal (sharp) interface between the two media, and
$\Delta\rho_{12}$ is the density contrast. Note that $\rho$ is obtained by the combination of the solid
surface and a step train of lipid bilayers, convolved with a function describing the positional
fluctuations and multiplied by a coverage function (see below). The critical momentum transfer or the
critical angle in $R_F$ is related to the density contrast by $q_z = 4\pi/\lambda ~\sin (\alpha_c)\simeq 4
\sqrt{\pi r_0 \Delta\rho_{12}}$, with $r_0$ denoting the classical electron radius. Absorption can be
accounted for by an imaginary component of the wave vector. The substrate/film interface is the only
relevant boundary for the $R_F$ factor due to the following reasons~: (i) in many cases, the beam impinges
on the sample through the water phase and there is almost no contrast between the film and water (for
X-rays) so no refraction takes place at the water/film interface; (ii) due to decreasing coverage of the
upper layers, the water/film interface is broad and not well defined, again leading to vanishing reflection
and refraction effects at this interface. We describe this feature by a monotonously decreasing coverage
function $c(N)$ with $c(1)=1$ and $c(N)=0$.

Using the linearity of the integrand in equation (\ref{eq.me}), the reflectivity amplitude can be split
into two parts $r_A (q_z) + r_B (q_z)~ e^{-\rmi q_z d_0}$. $r_A$ is due to the reflection from the density
increment at the substrate and $r_B$ represents the multilamellar bilayers. Taking $\sigma_s$ to denote the
rms roughness of the substrate, we get $r_A = \Delta\rho_{12} ~e^{-0.5 q_z^2 ~\sigma_s^2}$. Note that the
position of the first bilayer is shifted by $d_0$ with respect to the substrate (due to the presence of a
thin water layer). $r_B$ can be accounted for by specifying a structure factor $s(q_z)$
\begin{equation}
\label{struct} s(q_z)= \sum\limits^N_{n=1} ~e^{\rmi n q_z d }~ e^{-\frac{q_z^2 \sigma_n^2}{2}} ~c(N)~,
\end{equation}
with $\sigma_n$ the rms-fluctuation amplitude of the $n$-th bilayer, and the form factor~:
\begin{equation}
\label{form} \label{eq.fq} f(q_z)=\int^\frac{d}{2}_{-\frac{d}{2}} ~\frac{\partial\rho (z)}{\partial
z}e^{\rmi q_z z} \rmd z ~.
\end{equation}
The bilayer is parameterized in terms of its first $No$ Fourier coefficients $f_n$,
\begin{equation}
\label{eq.fc} \rho(z)=\rho_0 + \sum\limits^{No}_{n=1} f_n~v_n~\cos
\left [\frac{2\pi n ~z}{d} \right] ~.
\end{equation}
Note that due to the mirror plane symmetry of the bilayer, the phases $v_n=\pm 1$ are reduced to
positive/negative signs only, facilitating the phase problem enormously. In fact, the correct choice of
signs (up to orders $n=4$ or $n=5$) can in most cases be guessed from the knowledge of the basic bilayer
profile, if not deduced from the data, where sign changes are often accompanied by observable cusps in the
reflectivity curves. The integral of the form factor can be solved analytically, yielding~:
\begin{equation}
f(q_z)=\sum\limits^{No}_{n=1} ~f_n ~v_n \left[ \frac{\rmi ~ 8\pi^2
n^2 ~\sin(0.5 q_z d)}{{q_z}^2d^2-4{\pi}^2 n^2}\cos(n\pi) \right]
~.
\end{equation}
More details of this approach are discussed in \cite{Salditt:2002}. In practice, the range of the
reflectivity determines the number $No$ of orders which should be included. Note also that the
parameterization of $n$ Fourier coefficients can easily be changed by way of a linear transformation into a
parameterization of $n$ independent structural parameters of the bilayer, such as bilayer thickness
(headgroup peak-to-peak distance), density maximum in the headgroup region, density in the bilayer middle
plane, density of the water layer, etc. \cite{Salditt:2002}.

Figure \ref{fig:pure_OPPC} shows the reflectivity of multilamellar OPPC membranes, measured at the in-house
high resolution rotating anode reflectometer at T=45°C. The sample was prepared on a cleaned (111) Si wafer
by spreading from TFE \cite{Muenster:1999}. During the measurement, the sample was in the fluid
$L_{\alpha}$ phase, and the smectic parameter was $d \simeq 53 \mbox{\AA}$. After subtraction of the offset
scan (figure \ref{fig:pure_OPPC}(a)), the reflectivity curve was fitted to a model with 5 free Fourier
coefficients.

We treated the same data using the Fourier synthesis approach. While full $q$-range fitting gives the
$\rho(z)$ on an absolute scale \cite{Salditt:2002}, Fourier synthesis determines a relative profile
$\rho(z)=a\rho_0(z)+b$, with two open parameters $a$ and $b$. All results were compared to the electron
density profile as calculated from MD simulation data of Heller {\em et al.} \cite{Heller:1993}. The
scaling parameters $a$ and $b$ used are 1120 and 0.31 for Lorentz factor $q^{-2}_z$, and 440 and 0.322 for
Lorentz factor $q^{-1}_z$, respectively. The comparison clearly indicates that the Lorentz factor
$q_z^{-1}$ fails, while $q_z^{-2}$ gives good agreement. In this way, empirical correction terms can be
"calibrated".

\section{Thermal and static disorder, non-specular scattering}

In order to determine the density profiles $\rho (z)$ from $f(q_z)$, the effects of thermal and static
fluctuations ({\em e.g.} due to defects) in $s(q_z)$ have to be quantified. Thermal fluctuations are
dominant at full hydration, when the compressional modulus $B$ of the stack is small, and less important at
partial hydration, when $B$ is high. Furthermore, the solid surface effectively reduces thermal
fluctuations (in particular long range undulations), making it possible to get higher resolution profiles
$\rho(z)$ than in the bulk, even in fully hydrated states \cite{Salditt:1999}. To quantify the fluctuation
effects and to incorporate them in the reflectivity analysis, one can either treat the layers as discrete
\cite{Holyst:1991,Lei:1995,Romanov:2002}, or consider the stack as a continuous elastic medium, described
by the classical smectic energy \cite{deGennes:1993} $H / V = \frac{1}{2} ~B \left( \frac{\partial
u}{\partial z} \right)^2 + \frac{1}{2}~K (\nabla_{xy}^2 u)^2 $, where $u(x,y,z)$ is a continuum
displacement field of the membranes with respect to a perfect lattice. $B [\mbox{erg/cm}^3]$ and $K
[\mbox{erg/cm}]$ are the bulk moduli for compression and curvature, respectively. $K$ is related to the
bending modulus of a single membrane $K_s$ by $K=K_s/d$.

This latter approach is more tractable than the discrete one and yields similar results
\cite{Poniewierski:1993}. The boundary condition at the flat substrate can be taken into account either by
taking the associated surface tension to infinity \cite{Shalaginov:1993}, or directly by choosing for the
fluctuation modes an orthogonal set of eigenfunctions which vanish at the substrate \cite{deBoer:1999}.
Even in the continuous medium approach, the discrete nature of the stack must be taken into account by
limiting the number of distinct fluctuation modes to the number of bilayers, $N$ (this amounts to
restricting the summation to the first Brillouin zone), lest spurious divergences appear. The model yields
a complete description of the fluctuation spectrum, including the dependence of the correlation function on
$z$ and on the in-plane distance $r$. However, only the rms fluctuation amplitude $\sigma_n$ for each
bilayer is needed to describe the specular scattering. Its value is simply determined as
\cite{Constantin:2003}~:
\begin{equation}
\sigma_n = \eta \left ( \frac{d}{\pi} \right )^2 \sum_{n=1}^{N}
\frac{1}{2n-1} \, \, {\sin}^{2} \left ( \frac{2n-1}{2} \pi
\frac{n}{N} \right ) \, .
\end{equation}
\noindent where $\eta = \frac \pi2 \frac{k_BT} {d^2 \sqrt{KB} }$ is a dimensionless parameter first
introduced by Caill\'{e} \cite{Caille:1972}, which quantifies the importance of the fluctuations.

The most important kind of imperfection in lipid films is often the inhomogeneous coverage, {\em i.e.} the
distribution of the total number of bilayers $N$ on lateral length scales of several $\mu$m, deriving
either from the non-equilibrium deposition process or from an equilibrium dewetting instability
\cite{Gallice:2002}. The effect can be modelled by a coverage function for which a convenient analytical
form can be chosen as~:
\begin{equation}
c(n)= \left [ 1-\left ( \frac{n}{N} \right ) ^{\alpha} \right ] ^2~,
\end{equation}
\noindent where $\alpha$ is an empirical parameter controlling the degree of coverage. This is a convenient
method, but not a very precise one, insofar as the fluctuation spectrum is still calculated for a fixed
number of layers, $N$. A growing number of totally dewetted patches has been observed in thin
oligo-membrane films hydrated from water vapor. Another type of decreasing coverage is found in thick films
in excess water at high temperatures, where parts of the multilamellar stack unbind from the substrate or
from the underlying bilayers. Accompanying this effect, multilamellar vesicles can be observed in light
microscopy at the lipid/water interface \cite{Vogel:2000}. Other defects appearing in the lipid films are
the typical textural defects of the smectic phase, such as focal conics or oily streaks. They are expected
to give rise to isotropic Debye-Scherrer rings, as in figure 1(c). Finally, hydrophobic/hydrophilic
interactions may lead to defect structures terminating the bilayers at the edges. All of the above defects
are presumably accompanied by long range distortion fields, so that additional contributions to $\sigma_n$
aside from thermal fluctuations could be present. The density of defects can vary significantly depending
on the preparation scheme. To this end, we have strong indications that the type of solvent used in thick
spread lipid films is of paramount importance if a uniform film thickness is to be obtained, see figures
\ref{fig:microscopy} and \ref{fig:mix_PC_PE}.

\begin{figure}
\begin{center}
\epsfig{file=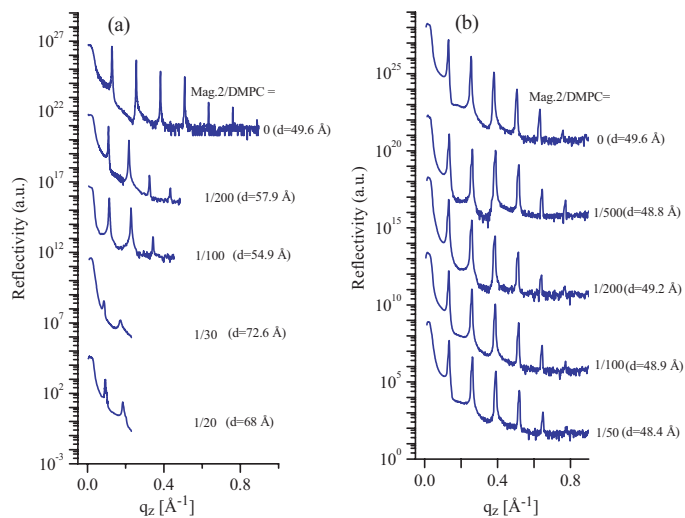,width=9cm}
\end{center}
\caption{\label{fig:compare_li_lu_data}Specular reflectivity curves of DMPC/magainin 2 at T=45°C (a) spread
from TFE on a (111) Si wafer, (b) spread from a 1:1 TFE:chloroform solution on glass. The curves are
shifted vertically for clarity, and run from low molar peptide concentration $P/L$ (top) to high $P/L$
(bottom).}
\end{figure}

While thermal fluctuations prevail at low osmotic pressure (high swelling), the dominant mechanism
determining the decay of higher order Bragg peaks at high osmotic pressure (low swelling) may be due to
static defects. This conclusion can be drawn from the fact that samples of nominally identical composition
and swelling may exhibit very different reflectivity curves, as illustrated by the comparison of two sample
series of magainin 2 in DMPC, one spread from TFE, see figure \ref{fig:compare_li_lu_data}(a), the other
from a TFE:Chloroform mixture, see figure \ref{fig:compare_li_lu_data}(b). Both series are shown as
measured before offset subtraction. Sample (a) was prepared on a cleaned (111) Si wafer and measured at the
in-house rotating anode reflectometer using Cu K$_{\alpha 1}$ radiation. Sample (b) was spread from the
mixed solvent on cleaned glass, and measured at the D4 bending magnet station, using 20 keV. Note that the
difference in photon energy leads to very different peak to tail ratios at the first Bragg peak, since the
interference with the substrate reflectivity amplitude is low for Cu K$_{\alpha 1}$ radiation due to
absorption. In figure \ref{fig:compare_li_lu_data}(a), the disorder in the multilamellar stack clearly
increases with peptide concentration, but in figure \ref{fig:compare_li_lu_data}(b) this effect is not
observed, possibly due to the different ($P/L$-dependent) sample quality (defect structure), see figure
\ref{fig:microscopy}. While for pure DMPC ($P/L=0$) six Bragg peak are observed in both cases, the
$P/L=0.01$ curve in (a) exhibits only three, the same curve in (b) five peaks. Thus, the Debye-Waller
factor must be significantly different. However, we must also note that the swelling was very different in
the two cases, such that part of the effect might be attributed to thermal fluctuations, which change with
the state of swelling.

The fact that TFE:Chloroform mixtures give particularly high quality samples is further illustrated by
figure \ref{fig:mix_PC_PE}, showing the reflectivity curves of samples composed of lipid mixtures of DMPC
and DMPE in a 1:1 molar ratio. The samples were prepared from 1:1 TFE:chloroform solutions, spread on (100)
Si wafers, and measured in the humidity chamber at a temperature $T\simeq51.6$°C at the D4 bending magnet
station of HASYLAB/DESY using photon energies of 20 keV. At this temperature and humidity, the samples are
probably in the gel phase, where thermal fluctuations are suppressed. Static defects seem to be minimized
by the preparation procedure, since an astonishingly high number of peaks (27 in pure DMPC:PE) can be
measured. Peptide addition slightly reduces the periodicity $d$, but does not affect the high number of
peaks.

\begin{figure}
\begin{center}
\epsfig{file=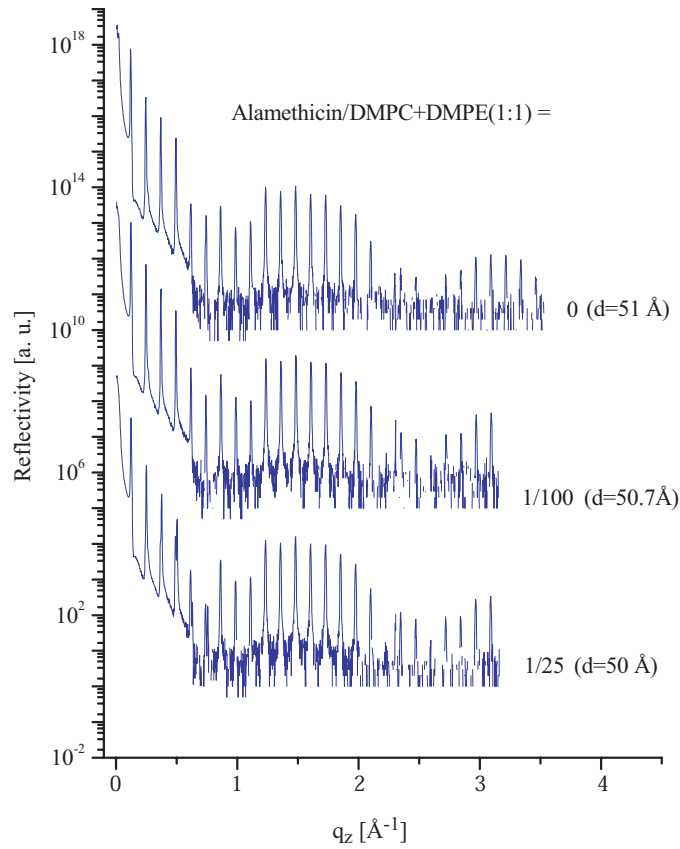,width=9cm}
\end{center}
\caption{\label{fig:mix_PC_PE} Offset-corrected reflectivity spectra of lipid mixture from DMPC and DMPE in
a 1:1 ratio at $T\simeq51.6$°C for different values of the alamethicin molar concentration $P/L$. The
samples were spread from a 1:1 TFE:chloroform mixture.}
\end{figure}

\section{Antibiotic peptides in lipid membrane model systems~: special approaches}

As mentioned in the introduction, different choices of sample preparation and measurement are possible, and
should be compared. Most of the studies discussed here were performed in humidity setups. However, in
biological systems the membranes are always in contact with a fluid phase; it is then very important to
study the biomimetic peptide-lipid systems in immersion, which has the additional advantage of allowing a
wider range of experimental conditions~: the use of osmotic stressors, already employed in the case of pure
lipid systems (see section 2) can control the hydration (and also the stability of the stack). Salt can be
used to screen the polar lipids and/or proteins. Finally, by stopflow experiments one could study the
interaction between peptides and various solutes, as well as the adsorption kinetics of the peptides
themselves onto the bilayers. Nevertheless, the presence of the solvent, which constitutes an additional
source for scattering, raises experimental problems~: the X-ray photons must be energetic enough to
penetrate about 1 cm of water; as for neutrons, it is sometimes more convenient to approach the sample
through the (crystalline) substrate.

\begin{figure}
\begin{center}
\epsfig{file=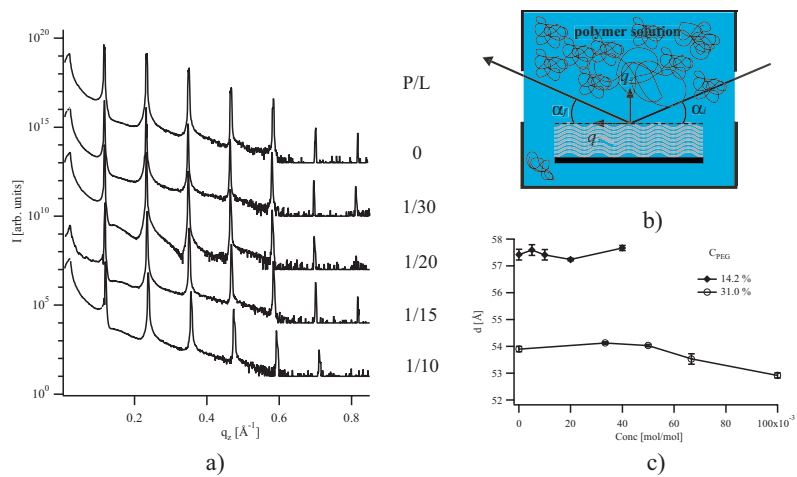,width=10.5cm}
\end{center}
\caption{\label{fig:Ala_DMPC} a) Reflectivity curves of DMPC multilayers containing Alamethicin immersed in
a $31 \%$ PEG solution (molecular weight $20 000$), with 100 mM NaCl, for different peptide/lipid ratios
$P/L$. b) Sketch of the experimental setup. c) Evolution of the periodicity $d$ with $P/L$ for two PEG
concentrations~: $14.2 \%$ (diamonds) and $31 \%$ (open dots).}
\end{figure}

As seen in the previous section, an important problem is the change in swelling state within one sample
series, see figure \ref{fig:compare_li_lu_data}. In principle, very important information is contained in
the slight changes of $d$, or --more accurately-- in the changes of water layer thickness $d_w$ and bilayer
thickness $d_{bl}$ as a function of $P/L$. However, there always remains a doubt whether the relative
humidity is really kept constant in the humidity cell for two consecutive samples and whether changes in
the swelling can really be attributed to changing interaction forces (e.g. electrostatic repulsion for
different $P/L$). By immersing the samples in calibrated solutions of osmotic stressors (polymer
solutions), this problem can also be circumvented. As an illustration, we present in figure
\ref{fig:Ala_DMPC}(a) reflectivity spectra of alamethicin-containing DMPC multilayers, in contact with a
$31 \%$ wt/vol PEG solution (molecular weight $20 000$), with 100 mM added NaCl, for different peptide
concentrations, measured on the ID1 beamline at the ESRF-Grenoble, with a photon energy of 19 keV. The
setup is represented in figure \ref{fig:Ala_DMPC}(b); the stainless steel chamber has kapton windows for
the incoming and the outgoing beam and is mounted on a heating stage for temperature control. In figure
\ref{fig:Ala_DMPC}(c) we show the repeat distance $d$ as a function of peptide concentration, for two
different PEG concentrations~: $14.2 \%$ (diamonds) and $31 \%$ (open dots). The error bars are obtained
from an average over three Bragg orders. A very slight decrease in $d$ takes place for $P/L > 1/20$ (only
reached for the series with $31 \%$ PEG solution shown in figure \ref{fig:Ala_DMPC}(a)). The physical
reason of the decrease is unclear at this point; as $d=d_{bl}+d_{w}$, a refined analysis of the form factor
is needed in order to discriminate between changes in the thickness of the lipid bilayer or the water
layer. Bilayer thinning at lower $P/L$ (and in different experimental conditions) was already reported (see
\cite{Chen:2003} and references therein).

\begin{figure}
\begin{center}
\epsfig{file=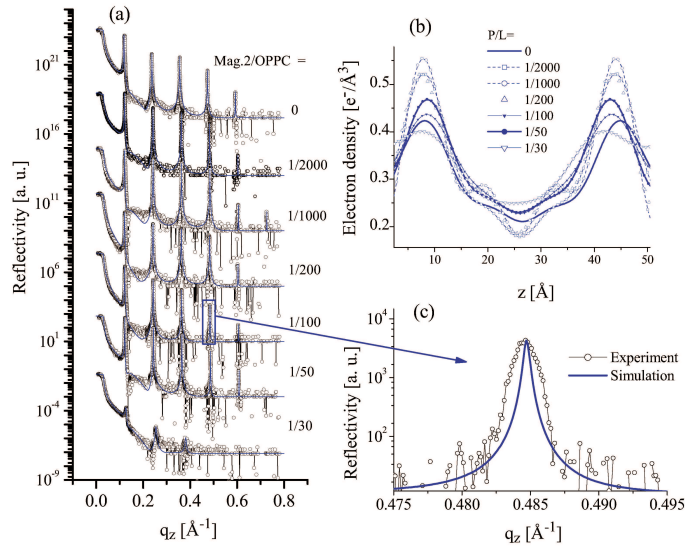,width=9cm}
\end{center}
\caption{\label{fig:OPPCmag_series}(a) Reflectivity of multilamellar samples of OPPC/magainin in the fluid
$L_{\alpha}$ phase at partial hydration ($T\simeq45$°C) with simulation (solid line), shifted for clarity.
(b) Electron density profile $\rho(z)$, obtained by fitting. (c) Zoom of a Bragg peak showing the
discrepancies due to instrumental resolution.}
\end{figure}

For comparison, a standard measurement in the humidity setup is shown in figure \ref{fig:OPPCmag_series}
for a series of magainin in OPPC as a function of $P/L$. The preparation and measurement setup is
equivalent to the pure OPPC sample shown in figure \ref{fig:pure_OPPC} (a).  The periodicity $d$ decreases
with $P/L$. After fitting the curves to the model described above (solid line) over the full $q_z$-range,
the profiles for $\rho(z)$ have been superimposed in (b). From these curves a monotonic decrease of the
bilayer thickness (defined as the head-head distance $d_{bl}$ from the maximum in $\rho(z)$) is inferred,
ranging from $d_{bl}= 36.2 \mbox{\AA}$ at $P/L=0$ to $d_{bl}= 34.2 \mbox{\AA}$ at $P/L=0.033$.
Surprisingly, the least-square fits give unrealistically high values for the electron density in the head
groups. This result has to be regarded as an artifact. The reason is probably due to the fact that, in
order to determine the absolute scale in $\rho(z)$ correctly, the lineshape and integrated peak intensity
of the Bragg peak has to be fitted correctly, which is not the case, see zoom shown in figure
\ref{fig:OPPCmag_series}(c). The problem here is that the instrumental resolution has not yet been taken
correctly into account. As a consequence, the profiles are flawed.

\section{Conclusions}

In conclusion, we have shown different experimental approaches for studies of peptide-lipid interaction by
X-ray scattering from solid-supported films. Sample preparation, measurement and data analysis are largely
analogous in the case of neutron reflectivity (data not shown here), which typically suffers from a smaller
accessible $q_z$-range due to less brilliant sources, but offers the advantage of contrast variation. More
generally, some aspects of sample preparation discussed here (in particular related to mosaicity and defect
density) may also be of interest for other techniques, both structural and spectroscopic.

While in principle able to distinguish different conformational states of macromolecules in and at the
bilayer, X-ray reflectivity analysis presents some important challenges, in particular related to the
correct model for thermal fluctuations and distortion fields by static defects. As a general strategy, it
is convenient to prepare sample series in several configurations and to measure them in different setups.
Simultaneous analysis of reflectivity curves obtained from thin oligo-membranes and thick multilamellar
stacks, and of curves obtained in humidity and immersion chambers, as well as specular and offset curves
provides a way to cross-check the results and to distinguish significant effects from artifacts. To this
end, the model should be improved by incorporating the resolution effects in the fitting program. In the
future, we will work towards a freely distributed software for reflectivity analysis for solid-supported
multilayers. One aspect which we have not addressed here but which is important for this goal is the proper
parameterisation of the macromolecules, {\em i.e.} the interpretation of the density profiles
\cite{Loesche:2002,WhiteHristova:2000}. Finally, synergies between advanced scattering methods and other
techniques (using oriented samples) such as solid-state NMR, infrared spectroscopy with site-directed
labels and optical dichroism should be exploited.

\ack

We are indebted to B. Bechinger for providing the magainin 2, and to Angel Mazuelas for help at the ID1
beamline. We thank HASYLAB/DESY and ESRF for providing  the necessary synchrotron beam time. Financial
support by the German Ministry of Research under grant BMBF 05KS1TSA/7 (Verbundforschung
Synchrotronstrahlung) is gratefully acknowledged. D.C. has been supported by a Marie Curie Fellowship of
the European Community programme \textit{Improving the Human Research Potential} under contract number
HPMF-CT-2002-01903.

\end{document}